# ARTICLE

# Patterning edge-like defects and tuning defective areas on the basal plane of ultra-large MoS$_2$ monolayers toward hydrogen evolution reaction

Bianca Rocha Florindo[a], Leonardo H. Hasimoto[a,b], Nicolli de Freitas[a], Graziâni Candiotto[e], Erika Nascimento Lima[f,g,h], Cláudia de Lourenço[a,c], Ana B. S. de Araujo[a,c], Carlos Ospina[a], Jefferson Bettini[a], Edson R. Leite[a], Renato S. Lima[a,b,c,d], Adalberto Fazzio[b,f], Rodrigo B. Capaz[a,e], Murilo Santhiago[a,b]*

The catalytic sites of MoS$_2$ monolayers towards hydrogen evolution are well known to be vacancies and edge-like defects. However, it is still very challenging to control the position, size, and defective areas on the basal plane of MoS$_2$ monolayers by most of defect-engineering routes. In this work, the fabrication of etched arrays on ultra-large supported and free-standing MoS$_2$ monolayers using focused ion beam (FIB) is reported for the first time. By tuning the Ga$^+$ ion dose, it is possible to confine defects near the etched edges or spread them over ultra-large areas on the basal plane. The electrocatalytic activity of the arrays toward hydrogen evolution reaction (HER) was measured by fabricating microelectrodes using a new method that preserves the catalytic sites. We demonstrate that the overpotential can be decreased up to 290 mV by assessing electrochemical activity only at the basal plane. High-resolution transmission electron microscopy images obtained on FIB patterned freestanding MoS$_2$ monolayers reveal the presence of amorphous regions and X-ray photoelectron spectroscopy indicates sulfur excess in these regions. Density-functional theory calculations provide identification of catalytic defect sites. Our results demonstrate a new rational control of amorphous-crystalline surface boundaries and future insight for defect optimization in MoS$_2$ monolayers.

## INTRODUCTION

As the world population grows, continuous efforts are expected to produce affordable, clean, and renewable energy that will be highly demanded for the next years.[1] One of the issues consists in replacing fossil fuels used in industrial processes with technologies that enable the use of renewable resources. Hydrogen (H$_2$) production, for instance, can play an important role in this field since it is a clean fuel and it has high energy density to supply future demands for renewable energy.[2,3] Water splitting is one of the most employed methods to produce H$_2$ and platinum is considered the best catalyst for this process. However, Pt is expensive and scarce, thus limiting the application of such catalyst for H$_2$ production.[4] Thereby, many efforts have been devoted to studying abundant and low-cost materials for hydrogen production.[5–7]

Transition metal dichalcogenides have been widely used in the energy conversion field due to their layered structure and ultrathin thickness.[8–13] These materials present a general formula MX$_2$, where M is a metal and X is a chalcogen. Among them, molybdenum disulfide (MoS$_2$) is one of the most studied materials for this purpose.[14–16] MoS$_2$ is low-cost, abundant, non-toxic, and it has been explored to substitute high-cost catalysts.[17] However, the catalytic activity of pristine MoS$_2$ toward hydrogen evolution mainly occurs at the edges of the 2D material.[18,19] In order to increase the number of catalytic sites, many methods have been reported to create defects on the basal plane, such as edges and sulfur vacancies.[20–22] Oxygen plasma,[23] argon plasma,[24] chemical treatment,[25] irradiation with He[26] and Ga$^+$ ions, to name a few, have been employed for this purpose. These strategies can generate mono and double sulfur vacancies or even holes in the 2D material. Oxygen plasma, for instance, can generate edge-like features on the basal plane. However, the precise tuning of size and position of edge-like defects, as well as the chemical selectivity of defects involved in the catalytic process, are very difficult to be controlled using the above-mentioned methods without the assistance of other techniques.

Different methods have been reported in the literature to introduce defects on the basal plane with spatial resolution. Irradiation with electron beam using transmission electron

[a] Brazilian Nanotechnology National Laboratory, Brazilian Center for Research in Energy and Materials, Campinas, São Paulo 13083-970, Brazil
[b] Federal University of ABC, Santo André, São Paulo 09210-580, Brazil
[c] Institute of Chemistry, University of Campinas, Campinas, São Paulo 13083-970, Brazil
[d] São Carlos Institute of Chemistry, University of São Paulo, São Carlos, São Paulo 09210-580, Brazil
[e] Instituto de Física, Universidade Federal do Rio de Janeiro, Rio de Janeiro, RJ 21941-972, Brazil
[f] Ilum School of Science, Brazilian Center for Research in Energy and Materials (CNPEM), Campinas, SP, Brazil
[g] Federal University of Rondonópolis, Rondonópolis, Mato Grosso 78736-900, Brazil
[h] Institute of Physics, Federal University of Mato Grosso, Cuiaba, Mato Grosso 78060-900, Brazil
* murilo.santhiago@lnnano.cnpem.br

Electronic Supplementary Information (ESI) available: [details of any supplementary information available should be included here]. See DOI: 10.1039/x0xx00000x

microscopy (TEM) is a remarkable example where single point defects and sulfur vacancies[27] can be created at specific areas of the free-standing monolayer. Lithography-based methods are used to open windows on the basal plane with microscale resolution that can be further submitted to plasma[28] or solution-processed routes to generate defects.[25] However, the removal of photoresist or polymethyl methacrylate (PMMA) can generate undesirable oxidation and contamination of the monolayer.[29,30] Ion beam with He[31] and scanning probe lithography[32] can generate defects with a high spatial resolution on monolayers. Such level of control is unique but time-consuming and incompatible with ultra-large areas. Thus, it is still challenging to establish a rational and controllable defect engineering route that allows precise control of position, size, and defective areas on the basal plane of monolayer $MoS_2$ that enables electrocatalytic studies.

Herein, we report the fabrication of large arrays of focused ion beam (FIB)-etched $MoS_2$ monolayers. Both suspended monolayers and monolayers supported on gold substrates were fabricated. Large $MoS_2$ monolayers were prepared directly on the substrates using a simple and fast method.[33] Next, gallium-FIB was used to etch $MoS_2$ and open reactive windows on the basal plane. The dose applied to etch a controlled depth on $MoS_2$ was calibrated using atomic force microscopy (AFM). The arrays were characterized using Raman spectroscopy, photoluminescence (PL), X-ray photoelectron spectroscopy (XPS), high-resolution transmission electron spectroscopy (HRTEM), and density functional theory (DFT) calculations, thus confirming the introduction of defects on the basal plane and identifying the chemical nature of such defects. By mapping the flakes using Raman and PL, we observed that defects reached longer distances from the etched edge when dose increases. HRTEM analysis in suspended monolayers reveals that defective regions become amorphous and XPS analysis indicate excess of sulfur in these regions, in variance with the majority of previous studies of defect-enhanced HER activity in TMDs, which focus on S-deficient defects. DFT calculations of H adsorption on model $MoS_2$ amorphous structures help to identify the chemical nature of S-rich defects most likely to be involved in the HER. The impact of each dose on the electrocatalytic activity toward hydrogen evolution reaction was measured by preparing microelectrodes.

## Experimental

### Electrode preparation

The electrodes were prepared by depositing 20 nm and 150 nm of Cr and Au, respectively, on glass slides using an electron beam system (AJA International). Before transferring the $MoS_2$ flakes the electrodes were cleaned using the following steps. First, a wet cleaning procedure using three solvents (acetone, isopropanol, and water) was performed. Next, the electrodes were submitted to using oxygen plasma (Dienner nano) for 10 min at a power of 30%, pressure of 0.3 mBar, and temperature of 100 °C. Then the bulk $MoS_2$ crystals (SPI supplies, PA, USA) were mechanically exfoliated using the "scotch-tape" method and transferred onto clean gold electrodes using a melted sucrose method. Sucrose (Synth, SP, Brazil) was melted at 250 °C and dropped on the tape containing previously bulk $MoS_2$ flakes. After solidification of sugar, the resulting solid piece was peeled off and positioned on the gold substrate and washed with heated (100 °C) de-ionized water leaving the $MoS_2$ flakes on the substrate.

### Electrochemical thinning and HER measurements

The $MoS_2$ monolayers were obtained by electrochemical thinning of bulk flakes using a three-electrode configuration in 0.5 M $H_2SO_4$ solution as the supporting electrolyte. The microfabricated Au electrodes were used as working electrodes, Ag/AgCl as reference electrodes, and a glassy carbon plate as counter electrode. The electrochemical thinning process was done using a PGSTAT-204 potentiostat model from AUTOLAB (Eco Chemie, Netherlands) interfaced with a computer and controlled by the NOVA 2.1 software. The process consists of two steps, first a chronoamperometry was proceeded by applying 1.5 V vs. Ag/AgCl for a certain time. Next, ten cyclic voltammetric cycles from -0.1 to 0.5 V at 10 mV s$^{-1}$ were performed in 0.5 M $H_2SO_4$, as reported.[33] Before assessing the electrocatalytic response the microelectrodes were cycled until a reproducible curve was attained. HER polarization curves were obtained at 5 mV s$^{-1}$ in a 0.5 M $H_2SO_4$ solution using the electrodes described above. The supporting electrolyte was purged with $N_2$ for 1 h before the experiments. The potential shift of Ag/AgCl electrode to RHE was calibrated as follows: $E(RHE) = E_{experimental} + 0.059 pH + 0.198$.

### TEM sample preparation

Free-standing $MoS_2$ samples were prepared by electrochemically thinning bulk $MoS_2$ samples on microhole-structured Ni-Au micromeshes. The micromeshes were prepared through photolithography and electrodeposition processes as reported.[33,34] The microfabricated micromesh is 40 μm thick and contains holes of 20 μm in diameter. The transfer process using sucrose was the same as described above on glass:Au electrodes, but additional cleaning steps were added, as follows. After sucrose transfer step, the sample was immersed in acetone, isopropanol, and water for 5 min in each solvent. Next, the samples were immersed in boiling water for 1 h and then exposed to water vapor for 1 h. After drying with $N_2$ gas the Ni-Au samples containing the bulk flakes were electrochemically thinned. Finally, the samples were cut and etching markers using FIB were added to identify the proper hole containing the free-standing monolayers (See SI – Fig. S1).

### FIB microfabrication

The etched windows were open on $MoS_2$ monolayers using a Helios Nanolab 660 containing a source of gallium ions. The ion beam dose was calibrated on thick bulk flakes. For the microfabrication of the arrays, 5 keV and 8 pA were selected in all experiments. Arrays of etched square windows of 5 x 5 μm spaced by 5 μm were made on monolayers attached on glass:Au samples using doses of 2.26×10$^{-12}$, 9.04×10$^{-12}$, 1.87×10$^{-11}$ and 3.65×10$^{-11}$ pC μm$^{-2}$, named D1 to D4, respectively. The same doses were applied on free-standing monolayers for HRTEM analysis. In this case, the size of the etched windows was 500 × 500 nm.

**Characterization**

Topographic images were acquired using atomic force microscopy (AFM, ParkSystems NX-10) in $N_2$ atmosphere applying the tapping mode of AFM with a FMR probe (NanoSensors), working with 75 kHz as nominal resonant frequency and 2.8 N/m as constant nominal springs. The acquired images were treated using Gwyddion software. Raman spectra were taken with a 532-nm laser and 50x objective lens (XploRA Plus Horiba), and the maps were acquired using a step of 0.75 μm in the range of 100 to 1100 cm$^{-1}$. Photoluminescence (PL) was also measured using XploRA Plus Horiba equipment with a 532-nm laser in the range of 535 to 800 nm using a 50x objective lens. X-ray photoelectron spectroscopy (XPS) was carried out with a Thermo Scientific Kα spectrometer (U.K.). Optical images were taken using a Zeiss Microscope Icc5 at different magnifications. Scanning electron microscopy (SEM) images were obtained using a Helios Nanolab 660 (Thermo Fischer Scientific). The free-standing $MoS_2$ samples were analyzed in a Double Aberration-Corrected Transmission Electron Microscope (TEM), Titan Cubed Themis, of the LNNano, at 80 kV, in Scanning Transmission Electron Microscopy (STEM) mode.

**Computational Methods**

The theoretical description of the $MoS_x$ structures was performed by first-principles calculations using density functional theory (DFT)[35,36] as implemented in the Vienna "ab initio" simulation package (VASP).[37] The exchange-correlation functional prescribed by Perdew, Burke, and Ernzerhof (PBE)[38] under the generalized gradient approximation (GGA) is used for the calculation of structural and electronic properties. The interactions between the valence electrons and the ionic cores are treated within the projector augmented wave (PAW) method.[39,40] The Brillouin zone (BZ) integrations are performed using a 6 × 6 × 1 Γ-centered Monkhorst-Pack sampling[41] for the structural optimization and 2 × 2 × 1 k points for ab initio molecular dynamics (AIMD) simulations. The NVT ensemble (the canonical ensemble, in which the amount of substance (N), volume (V), and temperature (T) are conserved) was used for AIMD simulations.

The electronic wave functions are expanded on a plane-wave basis with an energy cutoff of 400 eV. The $MoS_x$ supercells are on the x-y plane, where a 6 × 6 lateral periodicity was employed. To avoid interactions between the periodic images of the supercell, the systems are modeled using supercells repeated periodically on the x-y plane with a vacuum region of about 15 Å inserted along the z-direction. Spin-orbit coupling is included in all electronic structure calculations.

In order to predict the HER activity in the samples D3 and D4, we model the excess sulfur caused by FIB etching through random removal of Mo and S atoms on pristine $MoS_2$ monolayer until reaching the stoichiometry obtained in the experimental samples - $MoS_{2.2}$ and $MoS_{2.5}$. It is noteworthy that our goal is not to fully replicate the amorphous structure but to reveal the local structural changes that occur as a result of the FIB-etched.

The resulting $MoS_x$ structures were initially relaxed at 0 K to attain a local minimum. Next, to generate disorder, amorphous-like structures, we raised the temperature to 2000 K and held it for 10 picoseconds (ps). After that, each relaxed structure from the AIMD simulation was cooled from 2000 to 300 K for 1 ps. Finally, the thermal stability of all optimized structures was re-examined through a 10 ps AIMD simulation at 300 K.

The theoretical performances for HER of the $MoS_{2.2}$ and $MoS_{2.5}$ monolayers were estimated from the Gibbs free energy of adsorbed hydrogen ($\Delta G_{H^*}$) in all sulfur-sites (S-sites), using the single atom catalysis (SAC) approach.[42] The $\Delta G_{H^*}$ in this approach is defined as

$$\Delta G_{H^*} = E_{ads} + \Delta E_{ZPE} - T\Delta S_{H^*}, \quad (1)$$

where T is temperature, $\Delta E_{ZPE}$ and $\Delta S_{H^*}$ are the change in zero-point energy and entropy between atomic hydrogen adsorption and hydrogen in gas phase standard state (300K, 1 bar). Here, we approximate the entropy of hydrogen adsorption as $\Delta S_{H^*} \approx (1/2)S^0_{H_2}$, where $S^0_{H_2}$ is the entropy of gas phase $H_2$ at standard conditions. In this work, $\Delta E_{ZPE}$ was calculated for all S-sites of $MoS_x$, according to the following expression

$$\Delta E_{ZPE} = E_{ZPE(MoS_x+H)} - E_{ZPE(MoS_x)} - \frac{1}{2}E_{ZPE(H_2)}, \quad (2)$$

$E_{ZPE(MoS_x+H)}$ is the zero-point energy of one hydrogen atom adsorbed on $MoS_x$, $E_{ZPE(MoS_x)}$ is the zero-point energy of $MoS_x$ monolayer and $E_{ZPE(H_2)}$ is the zero-point energy of $H_2$ molecule in the gas phase. $E_{ads}$ is the hydrogen binding energy and can be defined as

$$E_{ads} = E_{(MoS_x+H)} - E_{(MoS_x)} - \frac{1}{2}E_{(H_2)}, \quad (3)$$

where $E_{(MoS_x+H)}$ is the total energy of $MoS_x$ monolayer with one hydrogen atom adsorbed, $E_{(MoS_x)}$ is the total energy of $MoS_x$ monolayer and $E_{(H_2)}$ is the total energy of $H_2$ molecule in the gas phase.

**RESULTS AND DISCUSSION**

Fig. 1 shows the schematic process to prepare ultra-large area monolayer $MoS_2$ on gold substrates and the FIB irradiation to pattern edges on the basal plane. In brief, we transferred bulk $MoS_2$ flakes onto gold substrates and applied an oxidation potential to thin the flakes down to a monolayer. This method enables the preparation of ultra-large area $MoS_2$ monolayers. The influence of time, oxidation potential, and thinning rate were systematically studied in our previous work.[33] Next, we created etched areas on the basal plane to verify the influence of edge-like defects created by Ga-irradiated samples. By using FIB, it is possible to prepare ultra-large arrays of defective regions with high spatial resolution. After irradiating the samples using FIB, we applied an insulating resin and exposed only the basal plane containing the etched windows to measure hydrogen evolution activity of the 2D material. We excluded the influence of the original edges of thinned flakes to account only for edge sites created by FIB, as will be shown ahead. The electrochemical activity of the etched windows toward hydrogen evolution was assessed using a three-electrode system.

First, we investigate the influence of Ga$^+$ FIB dose to reduce the thickness of $MoS_2$ flakes. The ion beam exposure dose (D) can be written as:

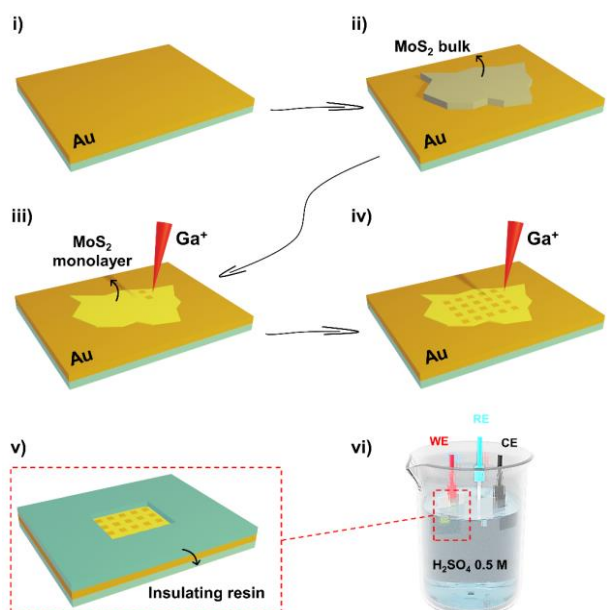

**Fig. 1.** Schematic figure illustrating the main fabrication steps: (i) thin Cr:Au film on glass substrates, ii) sucrose-mediated transfer of bulk MoS$_2$ flakes onto the Au electrodes, iii-iv) electrochemical thinning and FIB patterning of arrays of etched windows on the basal plane. v-vi) The array of etched windows was isolated and the electrochemical activity toward hydrogen evolution was measured using a three-electrode system.

$$D = \frac{I\,t}{q\,A} \qquad (4)$$

where I is the ion beam current, t is local exposure time, q is the ion charge, and A is the exposed area. Fig. 2a shows SEM images of bulk MoS$_2$ after the etching process by varying the dose in areas of 4 µm$^2$. The impact of each dose on the MoS$_2$ thickness was assessed by AFM, as illustrated in Fig. 2b. The images show the depth of the Ga-treated areas increases as a function of the dose applied on MoS$_2$. In addition, the depth vs. dose plot was found to be linear in the range studied, as shown in Fig. 2c. Thus, it is possible to tune the removal of approximately two MoS$_2$ layers from the surface with minimum impact on the microfabricated gold substrate. Next, the basal plane of electrochemically thinned MoS$_2$ was etched using FIB. The SEM image of the selected flake can be visualized in Fig. 2d. To pattern an array of etched windows (5 x 5 µm) on the basal plane, a dose of 2.26×10$^{-12}$ pC um$^{-2}$ (D1) was selected to remove approximately 1-2 layers. Additional results that confirm the etching process are shown in Fig. S2. Fig. 2e-g show images obtained at different magnifications of the etched areas from the entire array up to a single etched window. The etched lateral interface between Au and MoS$_2$ was studied using Raman and PL spectroscopies.

Raman spectroscopy is widely used to characterize MoS$_2$ since it is possible to determine many useful information, namely, structural phase, presence of defects and number of layers.[43,44] Fig. 2h shows single point Raman spectra acquired in the regions indicated by the inset numbers illustrated in Fig. 2g. The spectra obtained in position 1 (basal plane) shows the $E^1_{2g}$ (in-plane vibration) and $A_{1g}$ (out-of-plane) Raman active modes of 2H MoS$_2$ in 383 and 403 cm$^{-1}$, respectively. The frequency difference between the two modes was 19.7 cm$^{-1}$, thus showing the monolayer nature of MoS$_2$, in agreement with other studies.[45] At the etched interface (position 2), in addition to $E^1_{2g}$ and $A_{1g}$ modes, we observe a peak at 227 cm$^{-1}$. This new peak, LA mode, have been assigned to the presence of defects and disorder in MoS$_2$. The appearance of the LA peak has been observed by irradiating Mn$^+$, Ar$^+$ or Ga$^+$ on MoS$_2$ samples.[46–49] Lastly, the $E^1_{2g}$ and $A_{1g}$ were not detected in the inner region of the square, thus confirming that MoS$_2$ was fully etched.

PL experiments were also performed at the interface of the etched areas indicated in Fig. 2g. Monolayer MoS$_2$ shows a remarkable photoluminescence when compared to bulk MoS$_2$ due to changes in the electronic band gap from indirect (bulk) to direct (monolayer).[50] Fig. 2i shows the PL spectra obtained at different positions (1,2 and 3) of the interface. The spectrum obtained at the region of the basal plane (position 1), away from the etched area, shows a high-intensity peak located at 666 nm that can be attributed to exciton A. In addition, exciton B and negatively charged trions (A$^-$) can also be attributed to monolayer MoS$_2$ at 620 and 707 nm, as reported.[51] At position 2, a drastic decrease of PL intensity was observed due to the presence of defects, in consistency with the Raman results. It was reported that the increased presence of defects significantly quenches PL signal of MoS$_2$, generating more mid-gap defect states which ultimately results in an increased number of nonradiative decay channels.[52]

The impact of FIB on the Au-MoS$_2$ substrate was then investigated by Raman and PL maps to obtain a broader view of the surface changes due to introduction of defects. We changed the dose from 2.26×10$^{-12}$ to 3.65×10$^{-11}$ pC µm$^{-2}$, which represents an etching depth from 1.5 nm up to 15 nm, respectively. Fig. 3 shows the Raman and PL maps for each dose. In Fig. 3a we plot the Raman LA/$E^1_{2g}$ intensity ratio to show the impact of defects after each applied dose. For the dose D1, it is possible to observe that defects are mainly located near the etched edges. As the applied dose increases, the defective regions advance toward the basal plane. In the highest dose studied, the basal plane becomes almost completely defective, as indicated by the red regions in the Raman map for the dose D4. Since the presence of defects promotes a quenching in the PL signal, the PL maps illustrated in Fig. 3b provide complementary information regarding the above discussions. For dose D1, the PL signal is higher on the basal plane away from the etched regions. For higher doses, the defects are formed on large areas on the basal plane and the PL signal is quenched. In the dose D4, one can observe a very low and homogeneous PL intensity, thus confirming that defects spread over the entire basal plane. The combination of Raman and PL maps clearly shows the possibility to tune defective areas toward the basal plane by adjusting the applied dose. In addition, Fig. S3 shows no phase change to 1T' in the FIB patterned MoS$_2$ samples.

A common procedure to evaluate the electrochemical performance of the basal plane on monolayers is based on lithography fabrication routes. However, many unintentional modifications can occur on the surface of the 2D material due to organic solvent processing and photoresist residues that can hinder the electrocatalytic properties. For instance, the conductivity of MoS$_2$ can be drastically altered depending on the solvent used in the fabrication process.[29] In addition, photoresist contamination can be very critical in large-area MoS$_2$ monolayers and surface cleaning can promote additional defects.[30] For this reason, some attempts have been proposed to maintain the basal plane as pristine as possible using lithography-free methods.[29,53] This feature is particularly important when assessing the catalytic properties of the basal plane. Thus, we developed a two-step microfabrication process to avoid contamination on the patterned $MoS_2$ basal plane, as schematically illustrated in Fig. 4a and Fig. S4. Initially, an insulating resin is slowly approached in the desired area using a brush applicator to define the geometric area of the electrode. After, the solvent is left to evaporate, and the process is repeated to fully isolate the area.

Fig. 4b-c shows optical images of the array of etched windows before and after isolating the electrode, respectively. Thus, we avoid any residues from photolithographic process that may hinder the electrocatalytic activity of the 2D material. By applying the described two-step microfabrication process, the activity of microelectrodes with a high density of edges on the basal plane can be evaluated. We characterized the catalytic performance of the electrodes by obtaining polarization curves in 0.5 M $H_2SO_4$ solution, as can be viewed in Fig. 4d. As expected, the microelectrodes prepared on pristine large-area $MoS_2$ monolayer showed the highest overpotential ($\eta$) to drive HER due to the lack of active sites. After introducing defects, the overpotential necessary to reach 10 mA cm$^{-2}$ decreased up to 290 mV for ion dose D3 and then increased when a higher dose was applied. Fig. 4e also shows that the Tafel slope for dose D3 decreases when compared to pristine $MoS_2$, thus indicating enhanced HER activity. Stability studies are shown in Fig. S5.

High-resolution transmission electron microscopy is a remarkable characterization technique to investigate defects on the basal plane of 2D materials.[54] In the case of monolayers attached onto electrodes, the sample preparation involves a transfer step to TEM grids using PMMA. However, the transfer process introduces carbon contamination and creates unintentional defects on MoS$_2$ monolayers, making it difficult to establish a reliable correlation between defects and electrocatalytic activity. To circumvent these issues, we patterned etched windows using different Ga$^+$ doses directly on free-standing MoS$_2$ monolayers. First, Ni-Au flexible micromeshes were prepared using microfabrication processes[34]. Fig. 5a shows a 40 μm thick flexible micromesh. After sucrose transfer and electrochemical thinning, the micromesh was cut to fit into the TEM sample holder, as shown in Fig. 5b. The red arrows in Fig. 5c show ultra-large and free-standing monolayers to be patterned using FIB. By using electrochemical thinning, it is possible to minimize the presence of carbon contamination, as schematically shown in Fig. 5d. Residues of sucrose are removed when the upper MoS$_2$ layers are thinned. Thus, our method brings two remarkable features: (i) it minimizes carbon contamination and (ii) produces ultra-large free-standing monolayers. Such key advantages offer a large patio for defect engineering steps on monolayer MoS$_2$ using FIB and characterization using HRTEM.

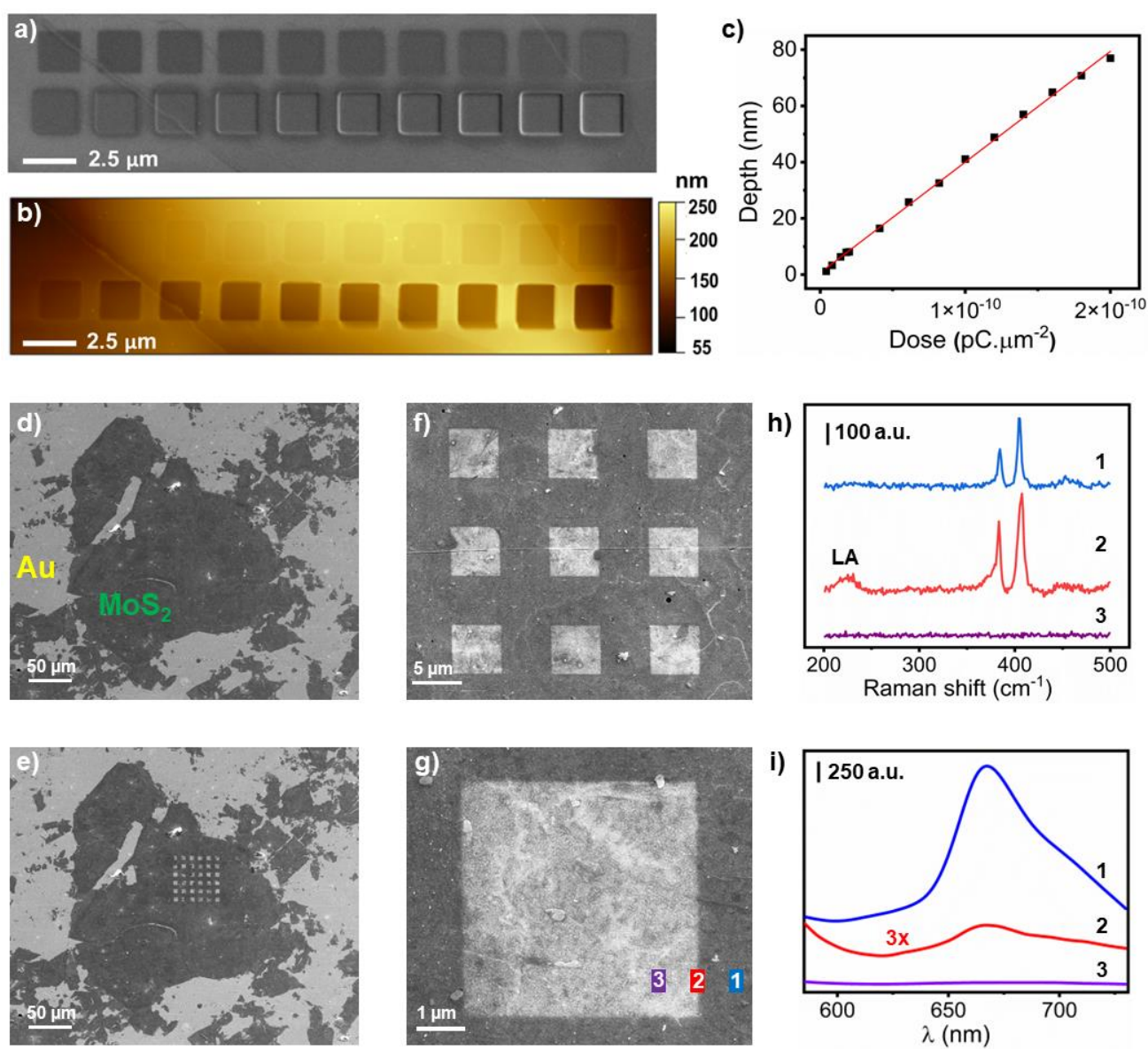

**Fig. 2.** a) SEM images of the etched areas on the basal plane of $MoS_2$. The applied dose increases from left to right on each row. b) AFM images of the etched areas. c) Graph of depth versus applied dose. d-g) SEM images of the etched areas on the basal plane of $MoS_2$. h) Single-point Raman spectra collected in the positions shown in the inset graph of Fig. 2g (positions 1,2,3). i) PL spectra obtained in the position shown in Fig. 2g.

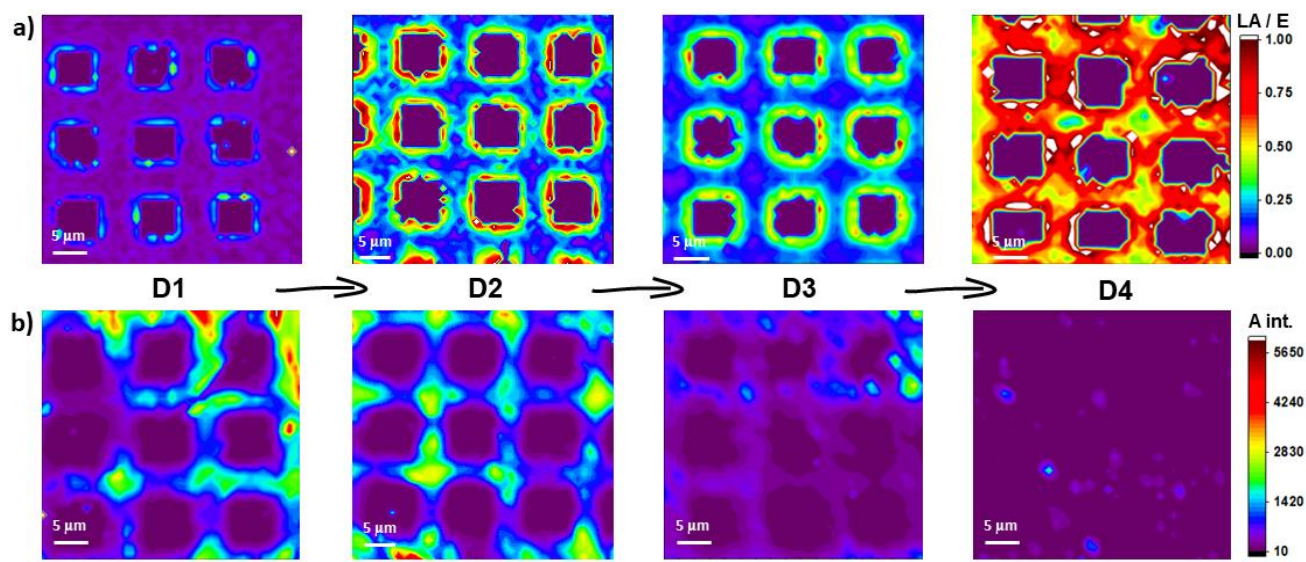

**Fig. 3.** a) Raman and b) Photoluminescence maps. Exposure doses D1 to D4 in this figure are $2.26\times10^{-12}$, $9.04\times10^{-12}$, $1.87\times10^{-11}$ and $3.65\times10^{-11}$ pC μm$^{-2}$, respectively.

Fig. 5e shows the array of etched windows 500 x 500 nm, on freestanding MoS$_2$, using the same doses D1 - D4 applied to obtain electrochemical information. To observe in greater detail the damage caused by the ion beam, we performed transmission electron microscopy measurements for all doses at different distances from the surface of the etched hole, as schematically shown in Fig. S6. Fig. 5f-j shows the HRTEM images for dose D1 and the positions for the damage study as a function of distance. The HRTEM images for the other doses investigated are shown in Fig. S7. From the Fourier transform of the high-resolution images, we obtained the pair distribution function, Fig. 5k, and, by the adjustment of a decreasing exponential function for the positive peaks, the length of structural coherence (LSC) was calculated to quantify the loss of coherence as a function of defects. As can be seen in Fig. 5k, the LSC value increases as a function of distance, indicating that damage is higher closer to the edge of the hole caused by the Ga$^+$ beam. Thus, our results indicated that the electrocatalytic activity at low overpotentials involves the presence of amorphous domain in the basal plane of MoS$_2$ monolayers.

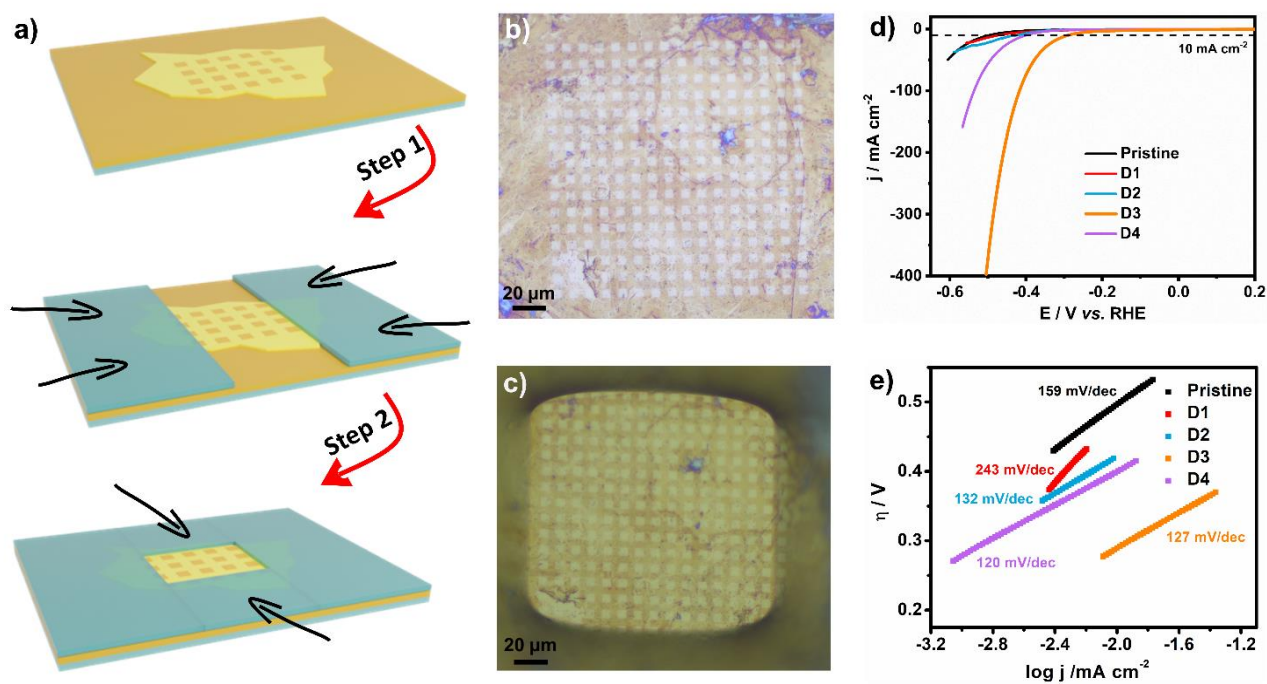

**Fig. 4.** a) Schematic representation of the two-step fabrication process. Stereomicroscope images of the array (b) before and (c) after applying the insulating resin. d) Polarization curves for all tested conditions. e) Tafel curves.

Amorphous MoS$_x$ materials have been demonstrated to present superior electrocatalytic activity when compared to their crystalline counterparts.[55,56] For instance, the shorter Mo-Mo bond distance assessed by X-ray absorption spectroscopy was attributed to facilitating H$^+$ adsorption on active sites and promoting higher HER activity.[57] In addition, apical and S-S terminated catalytic sites observed in electrodeposited amorphous MoS$_x$ thin films have been described to play an important role in the mechanism of HER.[56] Our XPS measurements on the basal plane identified similar surface chemistry observed in amorphous MoS$_x$ (Fig. S8). The spatial control of surface amorphization reported in this work provides unprecedented insight regarding the catalytic activity of amorphous-crystalline surface boundaries in monolayer $MoS_x$ toward HER. Our XPS spectra for samples D2, D3, and D4 indicate x = 2.2, 2.5, and 2.2, respectively, *i.e.*, we find an excess of sulfur in our amorphous samples and the extra amount of sulfur seems to behave non-monotonically as the $Ga^+$ beam dose increases. We recall that HER activity also behaves non-monotonically, increasing from the pristine sample to D2 and D3, and then decreasing from D3 to D4. Therefore, the HER activity seems to correlate nicely with the excess sulfur in the amorphous samples.

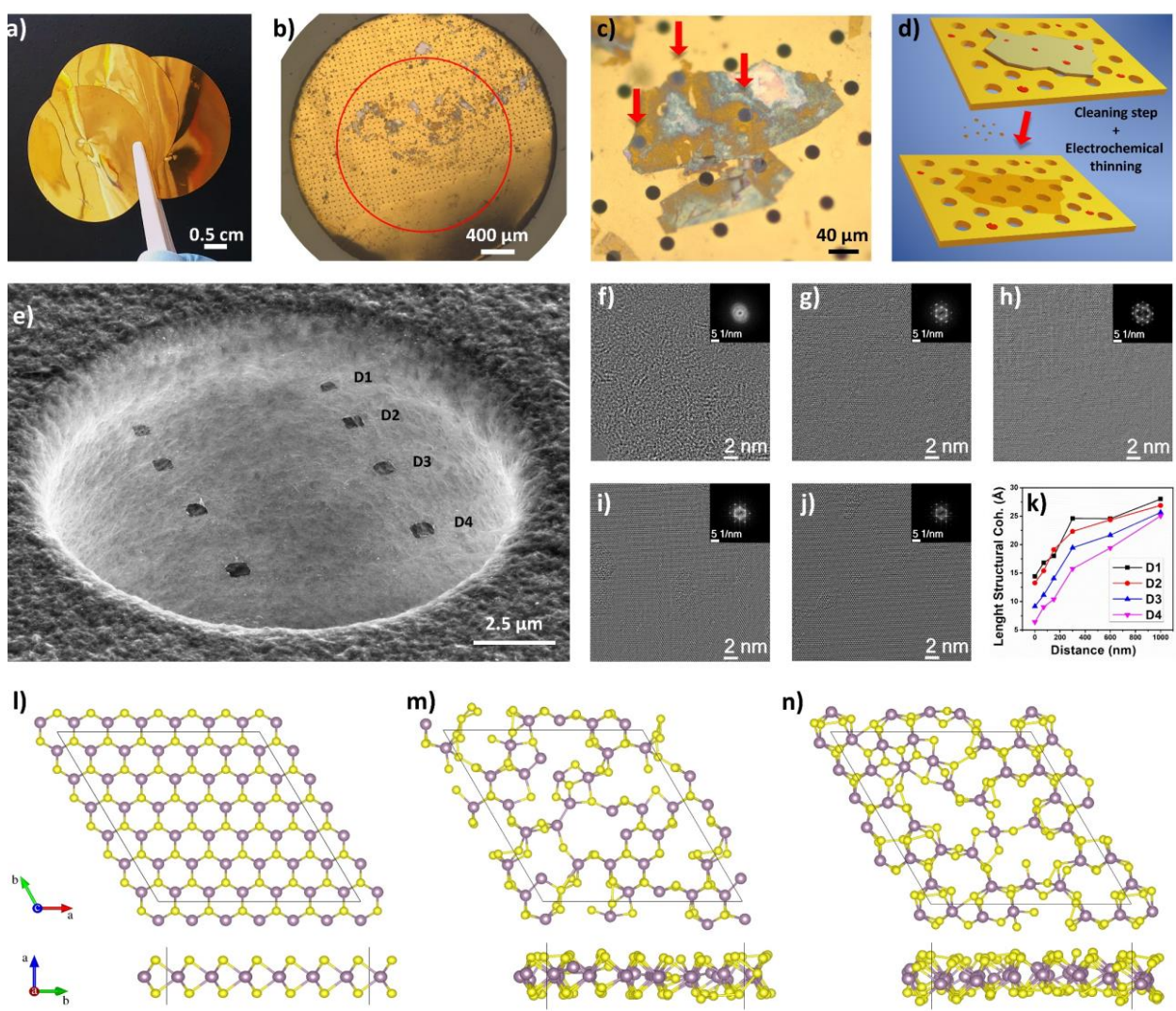

**Fig. 5.** a) Photo of the Ni-Au micromesh containing microholes of 20 μm in diameter. b) Stereomicroscope image of the micromesh after electrochemical thinning of bulk MoS$_2$ flakes. The micromesh was cut (d = 3mm) to fit into the TEM sample holder. c) High-magnification stereomicroscope image showing the free-standing regions indicated by the red arrows. d) Schematic illustration showing the thinning process and removal of carbon residues. e) Secondary electron image by SEM of one of the micromesh microholes, showing a set of etched windows applying four different exposure doses (D1-D4) on a free-standing MoS$_2$ monolayer. Different regions from the interface of the D1-nanohole where high-resolution images were obtained, starting from the interface, as follows: f) 0, g) 60, h) 150, i) 300, and j) 600 nm. k) Length structural coherence obtained for all doses studied (D1 – D4) as a function of the distance from the etched surface. Exposure doses D1 to D4 in this figure are 2.26×10$^{-12}$, 9.04×10$^{-12}$, 1.87×10$^{-11}$ and 3.65×10$^{-11}$ pC μm$^{-2}$, respectively. Top and side view of the optimized geometry of a 6x6 supercell for (l) pristine MoS$_2$, (m) MoS$_{2.2}$ and (n) MoS$_{2.5}$ monolayer. The purple and yellow colors correspond to the atoms Mo and S, respectively. The a, b, and c axes of the supercells employed in this work are indicated in the bottom.

To understand these results, theoretical studies based on DFT were conducted. For this analysis, we built a pristine 2H-$MoS_2$ (as a reference) and two amorphous (or amorphous-like) $MoS_x$ monolayers with x = 2.2 and 2.5 (see Fig. 5l, 5m, and 5n). The detailed methodology and step-by-step procedure for building the amorphous $MoS_x$ monolayers are described in Support Information.

It is known that for an ideal catalyst, the Gibbs free energy of adsorption (ΔG) should be approximately zero (ΔG ≈ 0), since η is related to ΔG by η=|ΔG|/e,[58] where e is the elementary charge. Thus, we address HER activity for $MoS_x$ monolayers by calculating the Gibbs free energy of adsorbed hydrogen ($ΔG_{H*}$) in the single atom catalyst (SAC) approach[42] for each sulfur site (S-site) in the model structures (see Fig. 5l-5n). For the pristine 2H-$MoS_2$ monolayer we find $|ΔG_{H*}|$ ≈ 1.93 eV, which indicates this structure does not exhibit good HER activity. This theoretical value for $ΔG_{H*}$ is confirmed by the experimental data (Fig. 4d) and it is in agreement with the literature.[59] For comparison, in the case of Pt we have $|\mathbf{ΔG_{H*}^{Pt}}|$ ≈ 0.09 eV.[59] However, the abundance of defects in the amorphous monolayers completely change their HER properties.[60] For the amorphous structure, we observe that about 10% of the S-sites shows $|ΔG_{H*}|$ ≤ 0.3 eV and 3.5% show $|ΔG_{H*}|$ ≤0.1 eV (in the same range of Pt catalysts) for the $MoS_{2.2}$ structures. For $MoS_{2.5}$, the number of active sites increases even further, where 21% of S-sites have $|ΔG_{H*}|$ ≤ 0.3 eV and 11.5% have $|ΔG_{H*}|$ ≤ 0.1 eV. Our results agree with the experimental findings of increased HER activity for amorphous structures of larger S content. In the case of $MoS_{2.5}$ and $MoS_{2.2}$ monolayers, the S-sites that has $|ΔG_{H*}|$ ≤ 0.1 eV are identified as being bridging disulfide ($S_2^{2-}$), that are widely reported as unsaturated sulfur edge.[61–65] Some of these sites are highlighted in the model structures of Fig. 5. Fig. S9 presents band structures for 2H-$MoS_2$, $MoS_{2.2}$, and $MoS_{2.5}$.The $ΔG_{H*}$ obtained for bridging $S_2^{2-}$ in this work are in agreement with previous works.[59] Our XPS analysis illustrated in Fig. S8, shows a considerable increase in the number of bridge sites in the D3 compared to other samples, thus confirming this interpretation.

Amorphization using FIB plays an important role to tune the electrocatalytic activity by introducing defects on pristine $MoS_2$ monolayer. The better electrocatalytic conditions were observed when amorphization was followed by a significant increase in the S/Mo ratio on the surface, as shown in Table S1. The electrocatalytic activity was improved when bridging $S_2^{2-}$/apical $S^{2-}$ increased on the surface, in agreement with other works.[66] Among these two sites, bridging $S_2^{2-}$ was theoretically found to have energy-favorable Gibbs free energy to drive HER.

## CONCLUSION

In this work we reported for the first time the precise tuning of edge-like defects with high spatial resolution on the basal plane of ultra-large $MoS_2$ monolayers using FIB. It was observed that amorphization in the neighbouring regions occurs during the process of opening etched windows on the basal plane. By tuning the exposed dose, it is possible to control the defective areas on the basal plane, from edge-confined amorphization to the complete amorphization of the entire $MoS_2$ basal plane. A combination of XPS measurements and DFT calculations reveal an excess of sulfur in the amorphous regions. Our route enables high spatial resolution on etching $MoS_2$ monolayers and unprecedently shows the controlled formation of crystalline-amorphous surface boundaries with tunable spacing on large-areas. By fabricating microelectrodes, we measured the electrocatalytic activity of the defective arrays and observed that the lower overpotential to drive HER depends on a proper $Ga^+$ dose. Again, the combination of XPS and DFT indicates that lower HER overpotential is associated to the increasing presence of bridging $S_2^{2-}$ in the amorphous regions as the amount of sulfur increases. HRTEM images, as a function of the distance from the etched edge, revealed that amorphization is higher near the edge and increases as a function of the dose. Our work paves the way for comprehending the influence of crystalline-amorphous domains in monolayer $MoS_2$ toward hydrogen evolution reaction.

## Author Contributions

B.R.F. contributed performing the transfer process, electrochemical thinning, and characterization studies. L.H.H. contributed with characterization studies and writing the manuscript. N.F., C.L., and A.B.S.A contributed with experimental procedures. J.B. and C.A.O. performed TEM analysis. G.C., E.N.L., A.F. and R.B.C. contributed with computational studies. G.C., E.N.L., R.B.C., E.R.L. and R.S.L. contributed with discussions. M.S. conceived the idea, designed the experiments, and wrote the manuscript.

## Conflicts of interest

There are no conflicts to declare.

## Acknowledgments

This work was supported by the Serrapilheira Institute (grant number: Serra-1912-31228) and FAPESP (grant number: 2022/00955-0), INCT - Materials Informatics and INCT - Carbon Nanomaterials. We thank Fabiano E. Montoro, Mariane P. Pereira, Davi H. Camargo, Otávio Berenguel, Carolina P. Torres, Carlos A. R. Costa, and Cleyton A. Biffe for their help in this work with instrumentation and laboratory facilities. G.C. gratefully acknowledges FAPERJ, grant number E-26/200.627/2022 and E-26/210.391/2022 (project Jovem Pesquisador Fluminense process number 271814) for financial support. The authors thanks Sistema Nacional de Processamento de Alto Desempenho (SINAPAD/SDUMONT) and Centro Nacional de Processamento de Alto Desempenho em São Paulo (CENAPAD-SP) for computational support. We also thank SisNano for the support.